\documentclass[12pt,a4paper]{article}
\usepackage{type1cm,amsmath,hangcaption,indentfirst}
\usepackage{amsfonts}
\usepackage{mathrsfs}
\usepackage{latexsym}
\usepackage{graphics}
\usepackage{sectsty}
\usepackage{exscale,latexsym}
\usepackage[dvips]{graphicx}
\usepackage{setspace}
\usepackage{multirow}
\usepackage{epsfig}
\usepackage{graphics}
\usepackage{graphicx,indentfirst}
\usepackage{psfrag}

\numberwithin{equation}{section}

\newcommand{\C}{\mathbb{C}}
\newcommand{\R}{\mathbb{R}}

\newcommand{\del}{\partial}
\newcommand{\str}{\text{str}}
\def\GL{{\rm GL(1$|$1)}}

\sectionfont{\normalsize\bf}
\subsectionfont{\bf\normalsize}
\def\be{\begin{equation}}
\def\ee{\end{equation}}

\def\C{{\mathbb C}}

\def\R{{\cal R}}

\def\nox{{\scriptstyle{\times \atop \times}}} 

\def\g{{\frak g}}

\def\boxit#1{\vbox{\hrule\hbox{\vrule\kern5pt
	 \vbox{\kern5pt#1\kern5pt}\kern5pt\vrule}\hrule}}

\def\Z{{\mathbb Z}}
\def\R{{\mathbb R}}

\def\noblackbox{\overfullrule=0pt}
\noblackbox

\def\L2{{\it Fun\/}\bigl(\text{\GL}\bigr)}

\newcommand{\ad}{\text{ad}}

\def\GL{{\rm GL(1$|$1)}}

\newcommand{\gzero}{\mathfrak{g}_{\bar{0}}}
\newcommand{\gone}{\mathfrak{g}_{\bar{1}}}

\newtheorem{thm}{Theorem}[section]

\newtheorem{definition}[thm]{Definition}

\begin{document}
\begin{titlepage}

 \renewcommand{\thefootnote}{\fnsymbol{footnote}}
\begin{flushright}
 \begin{tabular}{l}
 arXiv:1011.6424\\ 
 \end{tabular}
\end{flushright}

 \begin{center}

 \vskip 2.5 truecm

\noindent{\large \textbf{Yangian Superalgebras in Conformal Field Theory}}\\
\vspace{1.5cm}

\noindent{ Thomas Creutzig\footnote{E-mail: creutzig@physics.unc.edu}}
\bigskip

 \vskip .6 truecm
\centerline{\it Department of Physics and Astronomy, University of North Carolina,}
\centerline{\it Phillips Hall, CB 3255, Chapel Hill, NC 27599-3255, USA}

 \vskip .4 truecm

 \end{center}

 \vfill
\vskip 0.5 truecm

\begin{abstract}
Quantum Yangian symmetry in several sigma models with supergroup or supercoset as target is established.
Starting with a two-dimensional conformal field theory that has current symmetry of a Lie superalgebra with 
vanishing Killing form we construct
non-local charges and compute their properties. Yangian axioms are satisfied, except that the Serre relations only hold
for a subsector of the space of fields. 
Yangian symmetry implies that correlation functions of fields in this sector satisfy Ward identities. 
We then show that this symmetry is preserved by certain perturbations of the conformal field theory.  

\noindent The main example are sigma models of the supergroups PSL(N$|$N), OSP(2N+2$|$2N) and D(2,1;$\alpha$) away from the WZW point. 
Further there are the  OSP(2N+2$|$2N) Gross-Neveu models and current-current perturbations of ghost systems, both for the disc as world-sheet.
The latter we show to be equivalent to $\mathbb C\rm P^{N-1|N}$ sigma models,
 while the former are conjecturally dual to supersphere sigma models. 
\end{abstract}
\vfill
\vskip 0.5 truecm

\setcounter{footnote}{0}
\renewcommand{\thefootnote}{\arabic{footnote}}
\end{titlepage}

\newpage

\tableofcontents

\section{Introduction}
\label{sec:intoduction}

Due to its infinite-dimensional Virasoro symmetry two-dimensional conformal field theory belongs to the best understood
quantum field theories. Nonetheless, there exist many non-rational models which we cannot solve at the moment. 
Examples are sigma models on supergroups and supercosets with the property that the Killing form of the
 superalgebra of the global symmetry vanishes identically. These theories are argued to be conformal 
\cite{Bershadsky:1999hk, Kagan:2005wt, Babichenko:2006uc}, a recent article is \cite{Zarembo:2010sg}.
The problem is that generically these sigma models are only known to have Virasoro 
and finite-dimensional Lie superalgebra symmetry, 
which is in most cases not sufficient to determine physical quantities as spectra and correlation functions. 
It is thus crucial to find more symmetry and to learn how to use it.

The state of the art of sigma models on target superspace is that WZW models on supergroups are fairly well understood due to
their Kac-Moody symmetry, see 
\cite{Rozansky:1992td,Rozansky:1992rx,Schomerus:2005bf,Gotz:2006qp,Saleur:2006tf,Quella:2007hr,Hikida:2007sz,Creutzig:2010ne,Creutzig:2008an} 
for bulk and 
\cite{Creutzig:2008an,Creutzig:2009zz,Creutzig:2010zp,Creutzig:2008ag,Creutzig:2008ek,Creutzig:2007jy,Creutzig:2006wk} for boundary models.
The situation changes when Kac-Moody symmetry ceases to be present. Then perturbative computations starting from the WZW model
led to the computation of specific boundary spectra \cite{Quella:2007sg,Candu:2008yw,Candu:2009ep,Mitev:2008yt} 
and current correlation functions \cite{Benichou:2010ts,Benichou:2010rk,Ashok:2009jw,Ashok:2009xx,Konechny:2010nq}. In both cases the vanishing of
the Killing form simplified the perturbative treatment. Non-perturbatively one can single out some subsectors of the theory using the
global symmetry and compute correlation functions of fields in these sectors in simpler models \cite{Candu:2010yg}.

The motivation to study target superspace sigma models are its appearance in string theory and condensed matter physics. 
\newline String theory on AdS$_d$ $\times$ S$^d$ is described by superspace sigma models \cite{Metsaev:1998it,Berkovits:1999im,Berkovits:1999zq,Arutyunov:2008if,Stefanski:2008ik}. These theories are important as they are
conjectured to be dual to some gauge theories \cite{Maldacena:1997re}.
They are classical integrable \cite{Arutyunov:2008if,Bena:2003wd,Babichenko:2009dk}\footnote{In the AdS$_4\times\C$P$^4$ case classical integrability of the non-sigma model part of the string theory was found in \cite{Sorokin:2010wn}.}.
Quantum integrability is tightly connected with Yangian symmetry at the quantum level.
\newline In condensed matter physics in the case of the integer quantum Hall effect as well as for disordered systems 
psl(N$|$N) symmetry and
its central extension gl(N$|$N) are present \cite{Zirnbauer:1999ua, Guruswamy:1999hi}.  Boundary spectra were also computed perturbatively in this context \cite{Obuse:2008nc}.

An important quantity in the theory of ordinary Yangians is the eigenvalue of the quadratic Casimir element of the Lie subalgebra in the adjoint representation.
If the Killing form is degenerate then this operator acts non-invertible in the 
adjoint\footnote{This also applies to the superalgebras of type gl(N$|$N) where the Casimir acts non-zero but nilpotent in the adjoint.}.
 In this case the nature of the Yangian algebra is
expected to be rather unusual. Having a natural construction of the algebra as we provide it in this article should help
understanding these special Yangian superalgebras.
Reviews on ordinary Yangians are \cite{NMtwo,Bernard:1992ya}.

The present work is inspired by the following two developments. L\"uscher found non-local charges in massive two-dimensional quantum field theory \cite{Luscher:1977uq}, which were then identified as Yangian algebras by Bernard \cite{DB}.
These methods were extended to WZW models on compact Lie groups in \cite{Bernard2}. 
As conformal fields do not vanish at infinity the procedure depends on a point of reference $P$. 
\newline The second development is twistor string theory. It is argued to be dual to N=4 super Yang-Mills theory \cite{W, B}. Amplitudes in super Yang-Mills are computed to be Yangian invariant \cite{Drummond1,Drummond:2009fd}. The world-sheet theory of the open twistor string is essentially a free field representation of the psl(4$|$4) current algebra at level one. Using the methods of \cite{Bernard2} one can then establish Yangian symmetry \cite{Corn:2010uj}.
The procedure simplifies considerably due to the vanishing of the Killing form of psl(4$|$4), e.g. the dependence on the point of reference $P$ eventually vanishes. 
The main goal is to generalize this result and to establish Yangian symmetry in many conformal field theories based on superalgebras with vanishing Killing form and not necessarily possessing Kac-Moody current symmetry.

Our strategy is as follows. We consider Lie superalgebras whose Killing form vanishes identically and which possess a non-degenerate invariant bilinear form. This applies to the algebras of the groups PSL(N$|$N), OSP(2N+2$|$2N) and D(2,1;$\alpha$).
The first step is to take a conformal field theory with current symmetry of such a Lie superalgebra.
These currents are then used to construct non-local currents and charges, which depend on a point of reference $P$. 
Their properties are computed in complete analogy to the psl(4$|$4) level one example of the twistor string \cite{Corn:2010uj}.
The first result is that a special subsector of the conformal field theory carries a representation of the Yangian, given by the action of these non-local charges. Moreover this is a symmetry and gives interesting Ward identities. 
These identities are independent of the point of reference $P$. This leads us to define new charges which also provide a Yangian symmetry of this special subsector of the model.
The new charges allow us to investigate truly marginal perturbations that preserve the Yangian symmetry. 
We then list three examples. The main one are sigma models on the supergroups PSL(N$|$N), OSP(2N+2$|$2N) and D(2,1;$\alpha$).
Further, if we take the disc as world-sheet, we also have OSP(2N+2$|$2N) 
Gross-Neveu models which are conjectured to be dual to supersphere $\rm S^{2N+1|2N}$ sigma models. 
Finally again for the disc, we have perturbations of supersymmetric ghost-systems. 
We show that these are equivalent to sigma models on super projective space $\mathbb C\rm P^{N-1|N}$.
There are many more examples, as e.g. massive deformations, which we do not consider.

\section{Superalgebras}

We present the superalgebras that appear in the conformal field theories.

\subsection{Lie superalgebras}

We start with some properties of Lie superalgebras. They were first studied by Victor Kac \cite{Kac:1977em}.
A Lie superalgebra is defined as follows.
\begin{definition}
Let $\g$ be a $\mathbb{Z}_2$ graded algebra $\g=\gzero\oplus \gone$ with product 
$[\ \ ,\ \ ]:\g\times\g\rightarrow\g$ that respects the grading. The parity of a homogeneous element is denoted by
\begin{equation}
	\begin{split}
|X|\ = \ \Bigl\{ \begin{array}{cc}
	       \ 0&\qquad X \ \ \text{in}\ \ \gzero    \\
	       \ 1&\qquad X \ \ \text{in}\ \ \gone     \\
                        \end{array} \ .
		\end{split}
\end{equation}
Then $\g$ is a Lie superalgebra if it satisfies antisupersymmetry and graded Jacobi identity, i.e.
\begin{equation}
	\begin{split}
		0 \ &= \ [X,Y]+(-1)^{|X||Y|}[Y,X]  \qquad\text{and}\\[1mm]
	0\ &= \ (-1)^{|X||Z|}[X,[Y,Z]]+(-1)^{|Y||X|}[Y,[Z,X]]+(-1)^{|Z||Y|}[Z,[X,Y]]\ ,
\end{split}
\end{equation}
for all   $X,Y$ and $Z$ in $\mathfrak{g}$.

Further a bilinear form $B : \mathfrak{g} \times \mathfrak{g} \rightarrow \R (\text{resp.} \C)$
is called a consistent supersymmetric invariant bilinear form if
\begin{equation}
	\begin{split}
		B(X,Y) \ &= \ 0 \qquad\forall\, X \in \gzero \land \forall\, Y  \in \gone \\	
        	B(X,Y)-(-1)^{|X||Y|}B(Y,X ) \ &=\ 0 \qquad\forall\, X,Y\in \g \ \text{and}\\ 
	        B([X,Y],Z)-B(X,[Y,Z])\ &= \ 0 \qquad\forall\, X,Y,Z\in \g\ .\\  
\end{split}
\end{equation}	
\end{definition}
 Let $\g$ be a Lie superalgebra with non-degenerate invariant bilinear form $B$, $\{t^a\}$ a basis of $\g$ and ${f^{ab}}_c$ the structure constants, i.e. they satisfy
\begin{equation}
 [t^a,t^b] \ = \ {f^{ab}}_ct^c\, ,
\end{equation}
where the summation over the repeated index $c$ is understood.
Introduce the short-hand notation
\begin{equation}
(-1)^{ab} \ = \  (-1)^{|t^a||t^b|}\, .
\end{equation}
Then antisupersymmetry and Jacobi identity expressed in terms of structure constants are
\begin{equation}
\begin{split}
0\ &= \ {f^{ab}}_c - (-1)^{ab} {f^{ba}}_c \\ 
0\ &= \ (-1)^{ae} {f^{ab}}_c  {f^{de}}_b +   (-1)^{ad} {f^{db}}_c  {f^{ea}}_b +  (-1)^{de} {f^{eb}}_c  {f^{ad}}_b\, .\\
	\end{split}
\end{equation}
Let $\kappa^{ab}$ denote the invariant bilinear form $B(t^a  ,t^b  )$ of our Lie superalgebra $\g$. We choose the basis such that
\begin{equation}
   \kappa^{ab}\kappa^{bc}\ = \ (-1)^{a}\delta^{ac}\, ,
\end{equation}
which is always possible as can be seen by case by case analysis\footnote{Except for the exceptional Lie superalgebras it can be seen from the fundamental matrix representation.}.
Since our bilinear form is non-degenerate, we can define a dual basis $\{t_a\}$
\begin{equation}
     B(t^a,t_b)\ =\ \delta^a_b
\end{equation}
with dual metric $\kappa_{ab}$.
We raise and lower indices by using the metric, especially we need the formulae
\begin{equation}\label{eq:raising}
	\begin{split}
   {f^{ab}}_c \ &= \ {f^{abd}}\kappa_{cd}\qquad,\qquad
   {{f^{abc}}} \ = \  {f^{ab}}_d\kappa^{dc}\qquad, \qquad
   \kappa_{ab}\ = \ \kappa^{ab}\, .
   \end{split}
\end{equation}
The Lie superalgebras we are considering have vanishing Killing form, that is the supertrace in the adjoint representations vanishes, i.e.
\begin{equation}
\str(\ad(t^a)\ad(t^b)) \ = \  (-1)^d{f^{ac}}_d {f^{bd}}_c\ = \ 0\, . 
\end{equation}
Using \eqref{eq:raising}  we ca rewrite the vanishing of the Killing form in the following way 
\begin{equation}\label{eq:killing2}
  {f^{a}}_{bc} {f^{cb}}_d \ = \ (-1)^b  {f^{ab}}_c {f^{ec}}_b \kappa_{de} \ = \ 0\, .  
\end{equation}
This equation is important for the non-local charges we are going to construct.

\subsection{Yangian superalgebras}

Yangian algebras were first studied by Drindfel'd \cite{DRone,DRtwo}.

From now on let $\g$ be a simple Lie superalgebra with vanishing Killing form but with a non-degenerate invariant consistent bilinear form.
Then we define the Yangian superalgebra of $\g$.
\begin{definition}\label{def:ysa}
A Yangian superalgebra $(Y(\g), \Delta, \epsilon, S)$ 
is a $\Z_2$ graded algebra that is in addition graded by the non-negative integers
\begin{equation}
Y(\g) \ = \ \bigoplus_{n=0}^\infty \, Y_n
\end{equation}
such that the part of level zero is the Lie superalgebra $\g$ (with basis $\{t^a_0\}$), and the part of level one its adjoint representation (with basis $\{t^a_1\}$).
The higher level parts are then generated by the algebra product $[\ ,\ ]\,:\, Y(\g)\,\times\,Y(\g) \,\rightarrow\, Y(\g)$ subject to the 
constraint that the co-multiplication $\Delta\,:\, Y(\g)\,\rightarrow\,Y(\g)\,\otimes\,Y(\g)$ 
defined on the level zero and level one part as
\begin{equation}\label{eq:coproduct}
\begin{split}
\Delta(t^a_0)\ &= \ t^a_0\,\otimes\,1\,+\,1\,\otimes\,t^a_0 \\
\Delta(t^a_1)\ &= \ t^a_1\,\otimes\,1\,+\,1\,\otimes\,t^a_1\,+\, \alpha\, {f^a}_{bc}\,t^c_0\,\otimes\, t^b_0 \\
\end{split}
\end{equation}
is an algebra homomorphism.  In addition the co-multiplication has to be co-associative
\begin{equation}
(\Delta\,\otimes\, 1)\,\circ\, \Delta\ = \ (1\,\otimes\,\Delta)\,\circ\,\Delta\, .
\end{equation}
Further there is a co-unit $\epsilon\,:\, Y(\g)\,\rightarrow\,\R$
\begin{equation}
\epsilon(t^a_i) \ = \ 0 \qquad\text{for}\ i\,=\,0,1
\end{equation}
and an anti-pode $S\,:\, Y(\g)\,\rightarrow\,Y(\g)$
\begin{equation}
S(t^a_i) \ = \ -t^a_i \qquad\text{for}\ i\,=\,0,1
\end{equation}
\end{definition}
This defines a Hopf superalgebra. Note that the anti-pode axiom has such a simple form because of the vanishing of the Killing form.

The co-multiplication being a homomorphism restricts the algebra bracket of level one elements by some 
algebraic relations, the Serre relations. More precisely one can write the bracket as
\begin{equation}\label{eq:levelonebracket}
[t^a_1\,,\,t^b_1]\ = \ {f^{ab}}_c t^c_2 \, +\, X^{ab}\,.
\end{equation}
The Serre relations then fix $X^{ab}$ uniquely as follows (see \cite{NMtwo} for a similar argument).
Consider $u_{ab}$ satisfying 
\begin{equation}
u_{ab} \ = \ -(-1)^{ab}\,u_{ba} \quad\text{and}\quad u_{ab}\,{f^{ab}}_c \ = \ 0 
\end{equation}
this means that we can identify $u_{ab}$ with a closed element $u\,=\,u_{ab}\,t^a_0\,\wedge\,t^b_0$ in $\wedge^2\,\g\,\otimes\,\C$.
For homology of Lie superalgebras see
\cite{iohara}.
We define
\begin{equation}
Y^{ab} \ = \ (-1)^{de}\,{f^a}_{dc}\,{f^b}_{ge}\,{f^{dg}}_h\, t^c_0\,t^e_0\,t^h_0 \, .
\end{equation}
Vanishing of the Killing form implies that $Y^{ab}$ is totally supersymmetric, i.e. 
 we can view it as an element of Sym$^3(\g)$.
Requiring that $\Delta$ is a homomorphism leads to 
the Serre relations
\begin{equation}\label{eq:serre}
u_{ab}\,[t^a_1\,,\,t^b_1]\ = \ u_{ab}\,X^{ab}\ = \ u_{ab}\, \frac{\alpha^2}{3}\,Y^{ab}\,.
\end{equation}
If the second homology of $\wedge^*\,\g$ with coefficient in the trivial representation were zero, then these relation would be similar to
the standard Serre relations of Yangian algebras:
\begin{equation}
\begin{split}
-{f^{kl}}_i\,[t^i_1\,,\,t^j_1]\, + \, 
(-1)^{lj}{f^{kj}}_i\,[t^i_1\,,\,t^j_1]\, - \,
&(-1)^{k(l+j)}{f^{lj}}_i\,[t^i_1\,,\,t^k_1]\ = \\
 &= \alpha^2\,(-1)^{ih+gj}\,{f_{gid}}\,{f^{ld}}_{h}\,{f^{ki}}_{c}\,{f^{jg}}_e\, t^c_0\,t^h_0\,t^e_0 \, .
\end{split}
\end{equation}
This is the case for $\g\,\in\,\{osp(2N+2|2N), d(2,1;\alpha)\}$, but not for $psl(N|N)$ \cite{iohara}.

\subsection{Representations}\label{sec:rep}

A representation of the Yangian is also one of its Lie subalgebra. But not every Lie superalgebra representation can be lifted to the Yangian.
The simplest lift is the trivial one, i.e. let $\rho:\g  : \rightarrow \text{End}(V)$ be a representation of $\g$, then we define
$\tilde\rho: Y(\g) :\rightarrow \text{End}(V)$ by
\begin{equation}
\tilde\rho(t_0^a) \ = \ \rho(t_0^a)\qquad\text{and}\qquad\tilde\rho(t_1^a) \ = \ 0\, .
\end{equation}
Looking back to the Serre relations \eqref{eq:serre}, we see that this becomes a representation of the Yangian if and only if 
$\rho(Y^{ab})  =  0$.
It is not known in general when this condition is satisfied, however there are some criteria \cite{DRone, DNW2}.
These representations are important to us.

\subsection{Current algebra}

The conformal field theories we start to look at we require to possess currents $J^a(z)$ taking values in our Lie superalgebra $\g$ and satisfying the operator product expansion
\begin{equation}\label{eq:currentope}
J^a(z)J^b(w) \ \sim \ \frac{k\kappa^{ab}}{(z-w)^2} \ + \ \frac{{f^{ab}}_c \, J^c(w)}{(z-w)}\, .
\end{equation}
These currents have the operator valued Laurent expansion
\begin{equation}
J^a(z) \ = \ \sum_n \, J^a_n\, z^{-n-1}
\end{equation}
and the modes satisfy the relations of the affinization $\hat\g$ of $\g$
\begin{equation}
[J^a_n,J^b_m]\ = \ {f^{ab}}_c\, J^c_{n+m} \ + \ k\kappa^{ab} \delta_{n+m,0}\, ,
\end{equation}
where the central element takes the fixed value $k$, the level. Especially the zero modes form a copy of $\g$. 
The action of these modes on the vacuum state is
\begin{equation} 
J^a_n|0\rangle \ = \ 0 \ = \ \langle 0| J^a_{-n} \qquad\text{for} \ n\,\geq\, 0\, .
\end{equation} 
Important fields of the conformal field theory are the primary fields $\phi(z,\bar z)$. Let $\rho:\g\rightarrow\text{End}(V)$ be a representation of $\g$. Then to a vector $\phi$ in $V$, we associate the primary field $\phi(z,\bar z)$ and this field has the following operator product expansion with the currents
\begin{equation}
J^a(z)\phi(w,\bar w) \ \sim  \ \frac{\rho(t^a)\phi(w,\bar w)}{(z-w)} \, .
\end{equation}
The field $\phi(z,\bar z)$ is bosonic/fermionic if the vector $\phi$ is bosonic/fermionic and we denote its parity by $|\phi|$.
Finally, we define the subsector of primary fields that transform in representations that can be lifted trivially to representations of the Yangian
\begin{equation}\label{eq:goodsub}
\mathcal Y \ = \ \{\, \phi(z,\bar z)\, | \, \phi \ \text{in}\ V\ \text{and} \ \rho(Y^{ab})\,=\,0 \ \text{for} \ \rho : \g\rightarrow \text{End}(V)\,\}\, .  
\end{equation}
This terminates our preparations.

\section{Non-local Charges}
\label{sec:non-local}

Most of this section is in analogy of section four in \cite{Corn:2010uj}\footnote{Which is based on \cite{Bernard2}.}, which handled the special case of psl(4$|$4) current algebra at level one.
We start with a conformal field theory having a current algebra of $\g$, and $\g$ has vanishing Killing form as before. 
The zero modes of the currents form a copy of $\g$. They are the charges of level zero
\begin{equation}
Q^a_0 \ = \ J^a_0 \ = \ \frac{1}{2\pi i} \oint dz\, J^a(z)\, .
\end{equation}
The main goal of this section is to construct level one charges and to work out their properties. These charges correspond to non-local currents. 
Symmetries from non-local charges were considered in \cite{DR}.

We introduce the non-local field $\chi^a (z;P)$,
\begin{equation}\label{chi}
\chi^a(z;P)\ =\
\int^z_P dw J^a(w)
\end{equation}
and the non-local current
\begin{align}\label{eq:levelonecurrent}
Y^a(z;P) & =
{f^a}_{bc}  \int^z_P dw
J^c(w)\;J^b(z)
= {f^a}_{bc} \; \chi^c(z;P)\;J^b(z)\, .
\end{align}
Note that there is no normal ordering necessary, since
vanishing of the Killing form plus antisupersymmetry of the structure constants imply
\begin{equation}
{f^a}_{bc}\; J^c(w)\;J^b(z) \ \sim \ 0\, .
\end{equation}
We now investigate the corresponding charges and their action on primary fields.

\subsection{Co-algebra}

The co-algebra is determined by the action of the charges on a product of fields. 

The operator product with a primary $\phi(w, \bar w)$ is
\begin{equation}
\begin{split}
Y^a(z;P)  \; \phi(w, \bar w) \
&\sim \ {f^a}_{bc}\Big(\frac{ \nox\chi^c(w;P)\rho(t^b)\phi(w, \bar w)\nox}{(z-w)} + \\
 &\qquad\qquad-  \ln\Big(\frac{z-w}{P-w}\Big)\nox  J^c(z) \;\rho(t^b)\phi(w, \bar w)\nox \Big)\,.
\label{Yope}
\end{split}
\end{equation}
The action of the non-local charge $Q_1^a(P)$ corresponding to the current $Y(z;P)$ on the primary field $\phi(w, \bar w)$
is
\begin{equation}\label{eq:actionq1}
\begin{split}
Q^a_1(P)\big(\phi(w, \bar w)\big)
\ &=\  \oint_{\mathcal C_w} dz \,
Y^a(z;P)  \; \phi(w, \bar w)\  =
\ 2 \; {f^a}_{bc} \; \nox \chi^c(w;P)\;\rho(t^b)\phi(w, \bar w)\nox\,
,
\end{split}
\end{equation}
where the cut for the logarithm extends from $w$
passing through the point $P$, and for the contour $\mathcal C_w$ 
see Figure 1. The computation of the integral is then as in \cite{Corn:2010uj}.
\begin{figure}[!h]
	\label{fig:fig2}
\centering%
\psfrag{p}{\small $P$}
\psfrag{c}{\small $\mathcal C_w$}
\psfrag{w}{\small $w$}
\centering
\includegraphics[width=5cm]{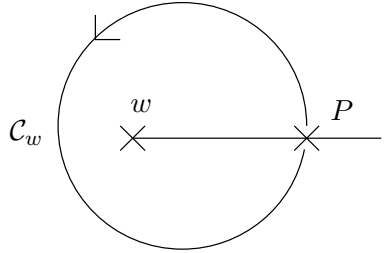}
\caption{\em The contour $\mathcal C_w$ starts
just above $P$, circles around $w$ and stops just below $P$.}
\end{figure}
The action of the non-local charge on two fields is 
\begin{equation}
\begin{split}
Q^a_1(P)\Big(\phi_1(w_1, \bar w_1)\,
\phi_2(w_2, \bar w_2)\Big)\ &=\
\oint_{\mathcal C_{w_1,w_2}}dz\, Y^a(z;P)\,\phi_1(w_1, \bar w_1)\,
\phi_2(w_2, \bar w_2)\\
&= \
Q^a_1(P)\Big(\phi_1(w_1, \bar w_1)\Big)\; \phi_2(w_2, \bar w_2)+ \\
&\qquad+ (-1)^{a|\phi_1|}\phi_1(w_1, \bar w_1)\; Q^a_1(P)\Big(\phi_2(w_2, \bar w_2)\Big)\, +\\
&\qquad - 2\pi i\;{f^a}_{bc}\;
 Q^c_0 \Big( \phi_1(w_1, \bar w_1) \Big)\;
Q^b_0 \Big( \phi_2(w_2, \bar w_2)\Big)\, .
\label{twoftran}
\end{split}
\end{equation}
For the contour $\mathcal C_{w_1,w_2}$ see Figure 2.
\begin{figure}[!h]
	\label{fig:fig3}
\centering%
\psfrag{P}{\small $P$}
\psfrag{xi1}{\small $w_1$}
\psfrag{xi2}{\small $w_2$}
\psfrag{cw}{\small $\mathcal C_{w_1,w_2}$}
\centering
\includegraphics[width=4cm]{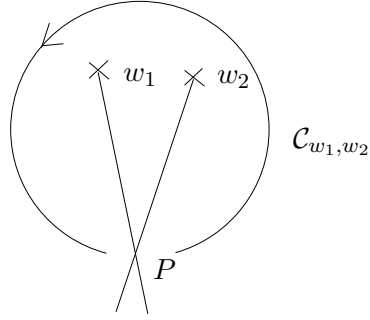}
\caption{\em The contour $\mathcal C_{w_1,w_2}$ starts at $P$ above both cuts
(cut from $w_1$ to $P$, and cut 
from $w_2$ to $P$), encircles both
$w_1, w_2$ and stops below both cuts at $P$.}
\end{figure}
Again the computation is exactly as in \cite{Corn:2010uj}.
The action on two fields can be written in terms of the co-product of the Yangian \eqref{eq:coproduct}
\begin{align}
\Delta (Q^a_1(P))\ &=\
Q^a_1(P) \otimes 1 + 1\otimes Q^a_1(P)- 2\pi i \; {f^a}_{bc}\; Q^c_0 \otimes
Q^b_0\,.
\label{ycoprod}\end{align}

\subsection{Algebra}

The algebra axiom is a straightforward computation. We compute
\begin{equation}
\begin{split}
Q^a_0\, Y^b(v;P) \ &= \ \frac{1}{2\pi i}\oint dz\, J^a(z)\, {f^b}_{cd}\,\int_P^vdv\, J^d(w)\,J^c(v) \\
&= \  {f^b}_{cd}\,\Big( {f^{ad}}_e\,\chi^e(v;P)\,J^c(v) +(-1)^{ad}{f^{ac}}_e\,\chi^d(v;P)\,J^e(v)\Big) \\
&= \ {f^{ab}}_c\, Y^c(v;P)\, .
\end{split}
\end{equation} 
The last equality follows from the Jacobi identity. 
Hence, the charges transform in the adjoint
\begin{equation}
[Q^a_0,Q^b_1(P)]\ = \ {f^{ab}}_c\, Q^c_1(P)\, .
\end{equation}

\subsection{Serre relations}

Recall that the co-multiplication is only well defined if the Serre relations \eqref{eq:serre} are satisfied.
If these relations fail to hold, the action of the level one Yangian charges on products of fields are preposterous.

We show that the left-hand side of the Serre relations vanish acting on one field.
This means that not the full space of fields of our conformal field theory forms a representation of the Yangian, but only the 
subspace $\mathcal Y$ \eqref{eq:goodsub} spanned by
those fields transforming in a representation that can be lifted trivially to a representation of the Yangian.

The left-hand side of the Serre relations \eqref{eq:serre} for the level one charges $Q_1(P)$ 
vanishes acting on a primary field $\phi(w, \bar w)$, if   
\begin{equation}\label{eq:leveloneserre}
[Q_1^a(P), Q_1^b(P)] \phi(w, \bar w) \ = \ {f^{ab}}_c X^c(w, \bar w;P)
\end{equation}
holds for some $X^c(w, \bar w;P)$. We show that the charges act indeed in this form.

We start with $Q_1(P)$ acting on $\chi(w;P)$.
We compute 
\begin{equation}
 \begin{split}
 {f^a}_{bc}J^c(w)J^b(z)J^d(v) \ \sim \  &{f^a}_{bc}\Big(\frac{k\kappa^{bd}J^c(w)}{(z-v)^2}+
\frac{(-1)^{bc}k\kappa^{cd}J^b(z)}{(w-v)^2} + \\
&+ \frac{{f^{bd}}_e \nox J^c(w)J^e(v)\nox }{(z-v)}+
\frac{(-1)^{bc}{f^{cd}}_e \nox J^b(z)J^e(v)\nox }{(w-v)}\Big)\, .
\end{split}
\end{equation}
Using the integrals in
 appendix \ref{app:integral}, we get
\begin{align}
 Q^a_1(P)(\chi^d(w;P)) \ = \ &{f^a}_{bc}{f^{bd}}_e\chi^c(w;P)\chi^e(w;P)-2{f^a}_{cb}k\kappa^{cd}\chi^b(w;P)\, .
\end{align}
This together with \eqref{eq:actionq1} implies
\begin{equation}
 \begin{split}
Q_1^a(P)(Q_1^b(P)(\phi(w, \bar w))) \ = \  &2{f^b}_{cd}Q_1^a\Big(\chi^d(w;P)\rho(t^c) \phi(w, \bar w)\Big) \\
= \ &2{f^b}_{cd}Q_1^a\Big(\chi^d(w;P)\Big)\rho(t^c) \phi(w, \bar w))\, +\\
&+ (-1)^{ad}2{f^b}_{cd}\chi^d(w;P)Q_1^a\Big(\rho(t^c) \phi(w, \bar w)\Big)\, .
\end{split}
\end{equation}
We combine this to
\begin{align}
 Q_1^a(P)(Q_1^b(P)(\phi(w, \bar w)))-(-1)^{ab}Q_1^b(P)(Q_1^a(P)(\phi(w, \bar w))) \ = \ & A^{ab} + B^{ab} 
\end{align}
with
\begin{equation}
 \begin{split}
A^{ab} \ &= \ 4{f^b}_{cd}{f^a}_{ef}(-1)^{ef}k\kappa^{fd}(\chi^e\rho(t^c)\phi)(w, \bar w;P)+\\
&\qquad -4(-1)^{ab}{f^a}_{cd}{f^b}_{ef}(-1)^{ef}k\kappa^{fd}(\chi^e\rho(t^c)\phi)(w, \bar w;P) \qquad\text{and} \\
 B^{ab} \ &= \ \Big({f^b}_{cd}{f^a}_{ef}{f^{ed}}_g +
 (-1)^{af}{f^b}_{df}{f^a}_{eg}{f^{ed}}_c +\cr
 \quad&\ -(-1)^{ab}{f^a}_{cd}{f^b}_{ef}{f^{ed}}_g -
 (-1)^{ab+bf}{f^a}_{df}{f^b}_{eg}{f^{ed}}_c\big) 
 2(\chi^f\chi^g\rho(t^c)\phi)(w, \bar w;P)\, .
\end{split}
\end{equation}
The Jacobi identity
implies
\begin{equation}
\begin{split}
 A^{ab} \ &= \ -4{f^{ab}}_d {f^d}_{ce}(\chi^e\rho(t^c) \phi)(w, \bar w;P)\, \\
 B^{ab} \ &= \ 2{f^{ab}}_{e}{f^{dc}}_{g}{f^{e}}_{df}(\chi^f\chi^g\rho(t^c)\phi)(w, \bar w;P)\, .
\end{split}
\end{equation}
We have thus shown \eqref{eq:leveloneserre},
and hence the left-hand side of the Serre relations vanishes acting on one primary field $\phi(w, \bar w)$.
This means that our charges $Q_0^a$ and $Q_1^b(P)$ form a representation of the Yangian if and only if 
the representation $\rho$ of the primary field can be lifted trivially to a representation of the Yangian
(recall subsection \ref{sec:rep}). 
Thus, we restrict to fields $\phi(z, \bar z)$ in $\mathcal Y$ .

\subsection{Co-unit}

Before we turn to Ward identities, we have to show that 
\begin{equation}
Q^a_0|\,0\,\rangle\ = \ Q^a_1(P)|\,0\,\rangle\ = \ 0\, .
\end{equation}
This means that the vacuum is a co-unit and implies Ward identities for these charges. 
Recall that the vacuum satisfies
\begin{equation}
J^a_n|\,0\,\rangle\ = \ 0\qquad\text{for}\ n\ \geq\ 0\, .
\end{equation}
Hence
\begin{equation}
Q^a_0|\,0\,\rangle\ = \ 0\, ,
\end{equation}
and further 
\begin{equation}
\begin{split}
Q^a_1(P)|\,0\,\rangle\ &= \ \frac{1}{2\pi i}\oint dz\, \int^z_P dw\, {f^a}_{bc} \, J^c(w)\, J^b(z)|\,0\,\rangle\\
&= \ \frac{1}{2\pi i}\oint dz\, \int^z_P dw\, {f^a}_{bc} \, \sum_{m,n}\, J^c_n\,w^{-n-1}\, J^b_m\, z^{-m-1}|\,0\,\rangle\\
&= \ 0\,.
\end{split}
\end{equation}
The last line follows from the vanishing of the Killing form plus supersymmetry of the invariant bilinear form.
In complete analogy one can show that 
\begin{equation}
\langle\,0\,|Q^a_0\ = \ \langle\,0\,|Q^a_1(P)\ = \ 0\, .
\end{equation}

\subsection{Ward identities}

In this section, we show that the non-local charges are indeed symmetries, that is they annihilate correlation functions.
But, we stress again that the action of the non-local charges on products of fields is only well defined if these fields
transform in representation that allow to be lifted trivially to a Yangian representation. Hence, the Ward identities only hold for correlation
functions containing these fields. Let $\phi_i(z_i, \bar z_i)$ be such fields. 

From last section, we get the Ward identities 
\begin{align}\label{Q1W1}
&\langle 0| Q^a_i  \phi_1(z_1,\bar z_1)\ldots \phi_n (z_n,\bar z_n) | 0\rangle \ = \ 0\,.
\end{align}
For the level zero charges this takes the form
\begin{equation}\label{eq:global}
\begin{split}
0\ &= \ 
 \sum_{i=1}^n \rho_i(t^a)\, \langle 0|  \phi_1(z_1,\bar z_1)\ldots \phi_n (z_n,\bar z_n) | 0\rangle\, .
\end{split}
\end{equation}
Here and in the following we use the notation
\begin{equation}
\begin{split}
&\rho_i(t^a)\, \langle 0|  \phi_1(z_1,\bar z_1)\ldots \phi_i(z_i, \bar z_i)\ldots \phi_n (z_n,\bar z_n) | 0\rangle \ = \\ 
&\qquad\qquad\qquad\qquad = \ \langle 0|  \phi_1(z_1,\bar z_1)\ldots \rho_i(t^a)\phi_i(z_i, \bar z_i)\ldots \phi_n (z_n,\bar z_n) | 0\rangle
\, .
\end{split}
\end{equation}
Using co-multiplication \eqref{ycoprod} and also \eqref{eq:actionq1} the level-one charges annihilating the vacuum imply
\begin{equation}
\begin{split}
0 \ =\ &\langle 0| Q^a_1(P) \phi_1(z_1,\bar z_1)\ldots \phi_n (z_n,\bar z_n) | 0\rangle  \\
= \ & 2 {f^a}_{bc}\;
\sum_{{i<j}} \; 
\rho_i(t^c) \;\rho_j(t^b) \; \ln\Big(\frac{z_i-z_j}{z_j-z_i}\Big)\; \langle 0| \phi_1(z_1,\bar z_1)\ldots \phi_n (z_n,\bar z_n) | 0\rangle
 + \\
&- 2 {f^a}_{bc}\;
\sum_{{i<j}} \; 
\rho_i(t^c) \;
\rho_j(t^b) \ln\Big(\frac{P-z_j}{P-z_i}\Big)\; \langle 0| \phi_1(z_1,\bar z_1)\ldots \phi_n (z_n,\bar z_n) | 0\rangle + \\
& +  2\pi i\,
{f^a}_{bc}\;\sum_{{i<j}} \; 
\rho_i(t^c) \;
\rho_j(t^b) \; \langle 0|  \phi_1(z_1,\bar z_1)\ldots \phi_n (z_n,\bar z_n) | 0\rangle \, . 
\end{split}
\end{equation}
We choose the phase\footnote{This means we insert the fields $\phi_i$ at points $z_i$ such that the phase is \eqref{eq:phase}. As this is certainly possible for an open domain of n-copies of the world-sheet there is no restriction. Also note, that other phase choices would give either trivial identities or Ward identities which can be obtained from the Ward identities we get plus crossing symmetry.} 
\begin{align}\label{eq:phase}
\ln\frac{(z_i-z_j)}{(z_j-z_i)}\; = \ln(-1) = \pi i \, .
\end{align} 
Using ordinary global symmetry \eqref{eq:global} of the correlation functions, antisupersymmetry of the structure constants and
the vanishing of the Killing form, we get
\begin{equation}
\begin{split}
{f^a}_{bc}\;
\sum_{{i<j}} \; 
\rho_i(t^c) \;
\rho_j(t^b) \ln\Big(\frac{P-z_j}{P-z_i}\Big)\; \langle 0|  \phi_1(z_1,\bar z_1)\ldots \phi_n (z_n,\bar z_n) | 0\rangle \ &= \ 0 \, .
\end{split}
\end{equation}
Thus, we arrive at the remarkable consequent of the Ward identity
\begin{equation}\label{ysymts}
{f^a}_{bc}\;\sum_{{i<j}} \; 
\rho_i(t^c) \;
\rho_j(t^b) \; \langle 0| \phi_1(z_1,\bar z_1)\ldots \phi_n (z_n,\bar z_n) | 0\rangle \ = \ 0\, . 
\end{equation}
Note that this identity does not depend on $P$. For two-point functions this identity follows from the vanishing of the Killing form plus global symmetry, while for more than two field insertions this is a non-trivial new identity, e.g. three-point functions obey
\begin{equation}
{f^a}_{bc}\;
\rho_1(t^c) \;
\rho_2(t^b) \; \langle 0| \phi_1(z_1,\bar z_1)\;\phi_2 (z_2,\bar z_2)\;  \phi_3 (z_3,\bar z_3) | 0\rangle\ = \ 0\,,\\
\end{equation}
where we simplified the identity using global symmetry and the vanishing of the Killing form.

\subsection{Yangian symmetry}
\label{sec:yangian}

We now have obtained our first goal, namely constructing non-local charges end establishing Yangian symmetry for a subsector of the 
conformal field theory. 
In this section, we find Yangian charges that are independent of $P$, all we need is the Ward identity \eqref{ysymts}.
These new charges have the advantage that we can use them to study perturbations of our conformal field theory.

We keep the action of the level zero charges as before, the level one generators we define to act trivially
on a single field
\begin{equation}
Q^a_1\big(\phi(z, \bar z)\big) \ = \ 0\, ,
\end{equation}
and the action on products of fields is defined by the Yangian co-multiplication plus its co-associativity.
This is only consistent with the Serre relations, if the right-hand side of the Serre relations acts trivially 
on a single field. 
We thus continue to restrict to fields $\phi(z, \bar z)$ in $\mathcal Y$.
The crucial observation now is, that this representation of the Yangian is also a symmetry of correlation functions
\begin{equation}
\begin{split}
&\Delta^n \big(Q_1^a \big) \; \; \langle 0| \phi_1(z_1,\bar z_1)\ldots \phi_n (z_n,\bar z_n) | 0\rangle \ = \\ 
&\qquad\qquad\qquad = \ {f^a}_{bc}\;\sum_{{i<j}} \; 
\rho_i(t^c) \;
\rho_j(t^b) \; \langle 0| \phi_1(z_1,\bar z_1)\ldots \phi_n (z_n,\bar z_n) | 0\rangle \ = \ 0\, . 
\end{split}
\end{equation}
The last equality is \eqref{ysymts}.

\subsection{Perturbations}

So far, we have found Yangian symmetry for a subsector of conformal field theories that already have affine Lie superalgebra symmetry.
Now, we turn to theories without Kac-Moody symmetry.

Conformal field theories with those Lie superalgebras as current symmetry which have zero Killing form possess interesting
deformations. We now introduce perturbations that preserve the symmetry we just found, i.e. they preserve global Lie superalgebra symmetry but
also the Ward identities \eqref{ysymts} will still hold and hence the subsector $\mathcal Y$ of fields will still have Yangian as symmetry.

Let $\mathcal O(z,\bar z)$ be a truly marginal operator that transforms in the trivial representation of the Lie superalgebra, i.e.
\begin{equation}
Q_0^a \mathcal O(z,\bar z) \ = \ \rho_{\mathcal O}(t^a)  \mathcal O(z,\bar z) \ = \ 0\, .
\end{equation}
Truly marginal means that the dimension of $\mathcal O(z,\bar z)$ is (1,1) and that a perturbation 
\begin{equation}
S_\lambda \ = \ S \,+\, \frac{\lambda}{2\pi} \int d^2z\,  \mathcal O(z,\bar z) 
\end{equation}
of our original conformal field theory with action $S$
by this field
is still conformally invariant.
Transforming in the trivial representation ensures that we keep global Lie supergroup symmetry, and hence fields still form a representation of
the finite-dimensional Lie superalgebra.

Let $\phi_i(z_i, \bar z_i)$ in $\mathcal Y$ as before.
A correlation function can be computed perturbatively as
\begin{equation}
\begin{split}
&\langle 0|  \phi_1(z_1,\bar z_1)
\ldots \phi_n (z_n, \bar z_n)  | 0\rangle_\lambda \ = \\
&\qquad\qquad\qquad  =\  \sum_m \langle 0|  \phi_1(z_1,\bar z_1)
\ldots \phi_n (z_n, \bar z_n) \frac{1}{m!}\Big(\frac{\lambda}{2\pi} \int d^2z\,  \mathcal O(z,\bar z)\Big)^m  | 0\rangle \, .
\end{split}
\end{equation}
The right-hand side is to be computed in the unperturbed theory. The trivial representation can obviously be lifted to the trivial
representation of the Yangian.
Since the perturbing field transforms by assumption in this representation, our Ward identities \eqref{ysymts} still hold 
in the perturbed theory
\begin{equation}
\begin{split}
0 \ &= \  \sum_m \Delta^{n+m} \big(Q_i^a \big)\langle 0|  \phi_1(z_1,\bar z_1)
\ldots \phi_n (z_n, \bar z_n)  \frac{1}{m!}\Big(\frac{\lambda}{2\pi} \int d^2z\,  \mathcal O(z,\bar z)\Big)^m  | 0\rangle\\ 
&= \  \sum_m \langle 0| \Big(\Delta^{n} \big(Q_i^a) \phi_1(z_1,\bar z_1)
\ldots \phi_n (z_n, \bar z_n)\Big)  \frac{1}{m!}\Big(\frac{\lambda}{2\pi} \int d^2z\,  \mathcal O(z,\bar z)\Big)^m  | 0\rangle \\
&= \
\Delta^n \big(Q_i^a \big) \; \langle 0|  \phi_1(z_1,\bar z_1)
\ldots \phi_n (z_n, \bar z_n)  | 0\rangle_\lambda  \\ 
\end{split}
\end{equation}
for $i=0,1$ and all $a$. 
But this says that the perturbed theory still satisfies
the nice Ward identity
\begin{equation}\label{eq:ward}
\begin{split}
\Delta^n \big(Q_1^a \big) \; \langle 0|  V_1(z_1)
V_2 (z_2)\ldots V_n (z_n)  | 0\rangle_\lambda \ &= \\ 
2\pi i{f^a}_{bc}\;\sum_{{i<j}} \; 
\rho_i(t^c) \;
\rho_j(t^b) \; \langle 0|  V_1(z_1)
V_2 (z_2)\ldots V_n (z_n)  | 0\rangle_\lambda \ &= \ 0\, . 
\end{split}
\end{equation}
and hence those fields that were forming a Yangian representation before perturbation still do so. Further
because of the Ward identity it is still a Yangian symmetry.

\section{Examples}

We now list some examples.
The first one is the CFT of twistor string theory, which inspired our analysis.
The other ones are sigma models on Lie supergroups, orthosymplectic Gross-Neveu models and sigma models on super projective space.
In the latter two cases, the world-sheet needs to be the disc.

\subsection{Supergroup sigma models}

Let the Lie supergroup be one of PSL(N$|$N), OSP(2N+2$|$2N) and D(2,1:$\alpha$). 
Then the Killing form of the underlying Lie superalgebra vanishes identically. 
Now, let $g(z,\bar z)$ be a supergroup valued field. The 
supergroup sigma model action 
is
\begin{equation}
S_{k,f}[g] \ = \ \frac{1}{2\pi f^2}\int d^2z\ \str\big(g^{-1}\del g g^{-1}\bar\del g\big) \ - \ 
\frac{k}{12\pi}\int d^2z\ d^{-1}\str\big((g^{-1}d g)^3\big)
\end{equation}
For ordinary Lie groups and supergroups this model is only quantum conformal for the Wess-Zumino-Witten model, 
i.e. for $k=f^{-2}$. In our case they are conformal at any coupling and at the WZW point we have additional Lie superalgebra
current symmetry. We thus view the sigma model as a perturbation from the WZW model
\begin{equation}
S_{k,f}[g] \ = \ S_{WZW}[g] \ - \ \frac{\lambda}{2\pi}\int d^2z\ \str\big(g^{-1}\del g g^{-1}\bar\del g\big) \, ,
\end{equation}
where $\lambda = f^{-2}-k$. 
The principal chiral field $\mathcal O_{PC} \, = \, \str\big(g^{-1}\del g g^{-1}\bar\del g\big)$ satisfies
\begin{equation}
Q_0^a\, \mathcal O_{PC}(z,\bar z) \ = \ 0\, .
\end{equation}
Hence at any sigma model on these supergroups possesses the additional symmetry. 
Especially let us mention again that the Ward identities \eqref{eq:ward}  
hold at any point $\lambda$. 

Remark, that we could have chosen a perturbation by the Wess-Zumino term equally well as the total action is invariant under the global symmetry.

\subsection{OSP(2N+2$|$2N) Gross-Neveu models}

Orthosymplectic Gross-Neveu models are supersymmetric generalizations of Gross-Neveu models for orthogonal groups. 
In the case of ordinary Gross-Neveu models the O(2)-model is special, as it is quantum conformal at any coupling. 
The reason for this is that O(2) is abelian. In the supergroup case despite being non-abelian the OSP(2N+2$|$2N)
Gross-Neveu models are still quantum conformal. These theories are conjectured to be dual to sigma models on the superpheres S$^{2N+1|2N}$ \cite{Candu:2008yw,Mitev:2008yt}.
The OSP(2N+2$|$2N) Gross-Neveu model is a current-current perturbation of a free theory given by 2N+2 real dimension $1/2$ fermions plus
N bosonic $\beta\gamma-$systems also of dimension $1/2$, 
\begin{equation}
 \begin{split}
S_\lambda\ &= \ S  \ +\ \frac{\lambda}{2\pi}  \int d^2z\, J^a \kappa_{ba}\bar J^b \\
S \ &= \ \frac{1}{2\pi}\int d^2z\, \Big( \sum_{i=1}^{2N+2} \psi_i\bar\del\psi_i+\bar\psi_i\del\bar\psi_i \Big)+ \\
&\qquad\qquad\qquad+ \frac{1}{2}\Big( \sum_{a=1}^{N} \beta_a\bar\del\gamma_a-
\gamma_a\bar\del \beta_a+ \bar\beta_a\del\bar\gamma_a-\bar\gamma_a\del \bar\beta_a        \Big)
\, .
\end{split}
\end{equation}
At $\lambda=0$ the model has a holomorphic and a anti-holomorphic osp(2N+2$|$2N) current symmetry. 
The holomorphic o(2N+2) currents are generated by the dimension one currents $\psi_i\psi_j$, the sp(2N) currents are generated by
the $\beta_a\beta_b, \gamma_a\beta_b$ and $\gamma_a\gamma_b$ and the fermionic currents are of the form $\psi_i\beta_a$ and $\psi_i\gamma_a$. This realisation is the affine analogue of an oszillator realisation for the horizontal subsuperalgebra for which the vanishing of the right-hand side of the Serre relations is understood \cite{Bargheer:2010hn}.
The current-current perturbation field $\mathcal O_{GN} =J^a \kappa_{ba}\bar J^b $ transforms in the adjoint representation under both holomorphic and anti-holomorphic
currents. It is thus not an operator that preserves Yangian symmetry. 
The situation changes if we take the world-sheet to have a boundary, i.e. the disc or equivalently the upper half plane.
In this case, we have to demand gluing conditions for the fields at the boundary.
We require that 
\begin{equation}
\psi_i(z) \ = \ \bar\psi_i(\bar z) \quad,\quad  
\beta_a(z) \ = \ \bar\beta_a(\bar z)\quad\text{and}\quad 
\gamma_a(z) \ = \ \bar\gamma_a(\bar z)\quad\text{for}\ z\ = \ \bar z\, . 
\end{equation}
This implies that the currents satisfy 
trivial boundary conditions $J^a(z)=\bar J^a(\bar z)$ for $z=\bar z$. 
Now, we view the anti-holomorphic coordinates as holomorphic coordinates on the lower half plane. 
We can then analytically continue the currents $J^a(z)$ on the entire plane 
\begin{equation}
\begin{split}
J^a(z)\ = \ \Bigl\{ \begin{array}{ccc}
\ J^a(z)&\qquad z \ \text{in upper half plane}     \\
\ \bar J^a(z)&\qquad z \ \text{in lower half plane}      \\
\end{array} \ 
\end{split}
\end{equation}
Now, the holomorphic currents as well as the anti-holomorphic currents transform in the adjoint representation of the charges $Q^a_0$ corresponding to the currents
that are defined on the full plane. 
It follows that the current-current perturbation transforms trivially,
\begin{equation}
Q^a_0 J^b(z)\kappa_{cb}\bar J^c(\bar z) \ = \ \big({f^{ab}}_d\kappa_{cb}+(-1)^{ad}{f^{ab}}_c\kappa_{bd}\big)J^d(z)\bar J^c(\bar z)\ = \ 0\, .
\end{equation}
Hence, the boundary Gross-Neveu model also has a subsector with Yangian symmetry. 
Again especially the Ward identity \eqref{eq:ward} 
holds. 

\subsection{Ghost-systems and $\mathbb{C}\rm{P}^{N-1|N}$ sigma models}

Our next example is given by $N$ fermionic ghost systems, each of central charge $c=-2$, and $N$ bosonic ghost systems each of central charge $c=2$. 
Such a system has a gl(N$|$N) current symmetry. 
Denote the bosonic ghosts by $\beta_i,\gamma^i$, $i=1,...,N$ and the fermionic ones by $\beta_i,\gamma^i$, $i=N+1,...,2N$.
Then the action is 
\begin{equation}
\begin{split}
S_\lambda\ &=\ S \ - \ \frac{\lambda}{2\pi}\int d^2z\, \mathcal{O}_{\beta\gamma} \\
S\ &= \ \frac{1}{2\pi}\int d^2z\, \beta_i (\bar\del+\bar A)\gamma^i +  \bar\beta_i (\del+ A)\bar\gamma^i\\
\mathcal{O}_{\beta\gamma} \ &= \ \mathcal{O}_{\beta}\mathcal{O}_{\gamma}\quad , \quad 
\mathcal{O}_{\beta} \ = \ \sum_{i} \beta_i\bar\beta_i \quad, \quad
\mathcal{O}_{\gamma} \ = \ \sum_{i} \gamma_i\bar\gamma_i\, .
\end{split}
\end{equation}
The gauge field $A,\bar A$ ensures that $\beta_i\gamma^i$ (often called U(1)$_R$ current) acts as zero and thus there is only sl(N$|$N) symmetry.
Further the total central charge is $c=-2$. The superalgebra sl(N$|$N) has a central element $E$, restricting to those representations
that possess $E$-eigenvalue zero effectively reduces the symmetry to psl(N$|$N). 
The perturbation is a current-current perturbation corresponding to the quadratic Casimir.
Yangian symmetry is preserved in the boundary case by the same arguments as in the Gross-Neveu model.
Interestingly all vertex operators that only depend on the $\gamma_i$ and $\bar\gamma_i$ obey the criterion of \cite{DNW2} as argued in the case $N=4$ \cite{Corn:2010uj}, i.e.
they allow for a trivial lift to a Yangian representation.

We are looking for a geometric interpretation of this theory. Recently path integral arguments led to correspondences 
\cite{Hikida:2007sz, Hikida:2007tq,CHR2}
and dualities \cite{Hikida:2008pe, CHR1} involving sigma models on non-compact (super) target spaces. 
In our case such arguments turn out to be very simple.

We restrict to correlation functions of fields $V_i(z_i)=V_i(\gamma(z_i),\bar\gamma(z_i))$ only depending on the $\gamma^i, \bar\gamma^i$ as for example gluon vertex operators in the twistor string.
Then performing the path integral first for the $\beta_i,\bar\beta_i$  gives
\begin{align}
\langle V_1(z_1)...V_n(z_n)\rangle_\lambda \ &= \ 
\int d^2\gamma d^2\beta d^2A\,  V_1(z_1)...V_n(z_n)\,e^{-S_\lambda[\beta, \bar\beta, \gamma, \bar\gamma, A, \bar A]}\nonumber\\ 
&= \ 
\int d^2\gamma d\bar\beta d^2A\, V_1(z_1)...V_n(z_n)\,\delta(\lambda\bar\beta_i\mathcal O_\gamma+(\bar\del+\bar A)\gamma_i)\,e^{-S_\lambda[0, \bar\beta, \gamma, \bar\gamma, A, \bar A]}\nonumber\\ 
&= \ 
\int d^2\gamma  d^2A\, V_1(z_1)...V_n(z_n)\,e^{-S^1_\lambda[ \gamma, \bar\gamma, A, \bar A]} \\ \nonumber
\text{where} \ S_\lambda^1[ \gamma, \bar\gamma, A, \bar A] \ &= \ \frac{1}{2\pi\lambda}\int d^2z\, \frac{1}{\mathcal O_\gamma}\sum_i(\bar\del+\bar A)\gamma^i(\del+A)\bar\gamma^i  \, .
\end{align}
Further integrating the gauge fields and redefining
\begin{equation}
(z_i,\bar z_i)\ = \ \frac{1}{\mathcal O_\gamma} (\gamma_i,\bar\gamma_i)
\end{equation}
yields the action of the $\mathbb{C}\rm P^{N-1|N}$ sigma model with couplings
$\theta=\frac{2}{\lambda}$ and $g^2=\frac{\pi\lambda}{2}$ \cite{Candu:2009ep}.
In summary, the complete (not only a subsector) boundary $\mathbb{C}P^{N-1|N}$ sigma model at couplings
$\theta=\frac{2}{\lambda}$ and $g^2=\frac{\pi\lambda}{2}$ has Yangian symmetry.

\section{Outlook}
\label{sec:conclusion}

We have found Yangian symmetry in a subsector of many conformal field theories with global superalgebra symmetry. 
The underlying Lie superalgebra has the crucial property that its Killing form vanishes identically. 
Our derivation was divided into several steps. The starting point was a conformal field theory that possesses Kac-Moody current symmetry.
Non-local charges depending on a point $P$ were constructed in terms of these currents and their properties were computed. 
The crucial result was that correlation functions of primaries in the subsector $\mathcal Y$, that allows for trivial lifts to Yangian representation,
obey the Ward identity \eqref{eq:ward}. This identity  is independent of $P$. This observation led us to define new charges that are again a symmetry for the subsector $\mathcal Y$ due to the Ward identity \eqref{eq:ward}.
Then perturbations that preserve this Yangian but not the Kac-Moody symmetry were considered. Finally, we gave some examples. We also derived a first order formulation for $\mathbb C\rm P^{N-1|N}$ sigma models.

There are several open questions. 
\newline
The apparent problem is to understand which representation of the Lie superalgebra can be 
lifted trivially to a representation of the Yangian, i.e. to identify the subsector of the conformal field theory
having Yangian symmetry. 
Once knowing these representation one should search for solutions to the Ward identities and classify them. 
We hope that they constrain correlation functions severely.
\newline
The list of examples can certainly be extended, e.g. one could search for massive deformations, 
one can apply it to world-sheet supersymmetric supergroup WZW models \cite{Creutzig:2009fh}.
Also our method applies to the current-current deformations recently studied in \cite{Konechny:2010nq} in this case for the disc as world-sheet. 
Here it would be interesting to understand how the Yangian symmetry is preserved by current-current perturbations according to the quadratic Casimir for the sphere as world-sheet.

\subsection*{\bf Acknowledgement}

I am very grateful to Louise Dolan for regular discussions and helpful comments. I also would like
to thank Brendan McLellan for discussions on Yangian superalgebras and their representations, and Peter R\o nne, Till Bargheer, Dmitri Sorokin, Cosmas Zachos and Jan Plefka for their comments.

\appendix

\section{Useful integrals}\label{app:integral}

Let $f(z)$ be an analytic function inside and on the contour $\mathcal{C}_x$.
For us $f$ is either $J$ or $\chi$. Note, that the contour is not going around the point $P$.
Then \cite{Corn:2010uj, Bernard2} have the following integrals
\begin{equation}\label{eq:intbernard}
\begin{split}
 \frac{1}{2\pi i}\oint_{\mathcal{C}_x} dz\, f(z) \ln\Big(\frac{z-x}{P-x}\Big) \ &= \ \int^P_x dz\, f(z)\\
\frac{1}{2\pi i}\oint_{\mathcal{C}_x} dz\, f(z) \int^v_Pdv\,  \frac{g(v)}{(z-v)} \ &= \ 
 \int_P^x dz\, f(z) \, g(z)\\
 \frac{1}{2\pi i}\oint_{\mathcal{C}_x} dz\, f(z) \int_P^x dv\, \frac{1}{(z-v)^2} \ &= \ 
  f(x) \\
 \frac{1}{2\pi i}\oint_{\mathcal{C}_x} dz\, f(z) \int^z_Pdw\,\int_{P'}^x dv\, \frac{1}{(w-v)^2} \
&= \ \int^{P'}_x dv\, f(v)\\
 \oint_{\mathcal{C}_x} dz\, \int_P^z dw\, \int_{P'}^x dv\, \frac{g(z)f(w)}{(w-v)} \ &= \ 0\, .
\end{split}
\end{equation}


\providecommand{\bysame}{\leavevmode\hbox to3em{\hrulefill}\thinspace}
\providecommand{\MR}{\relax\ifhmode\unskip\space\fi MR }
\providecommand{\MRhref}[2]{%
  \href{http://www.ams.org/mathscinet-getitem?mr=#1}{#2}
}
\providecommand{\href}[2]{#2}

\end{document}